\documentclass[11pt]{article}

\usepackage{amsxtra,amssymb,amsthm,amsmath,latexsym}
\usepackage{graphicx}
\usepackage{epsfig}
\usepackage{afterpage}

\newcommand{\R}{{\mathbb R}}

\newcommand{\bee}{\begin{equation*}}
\newcommand{\eee}{\end{equation*}}
\newcommand{\be}{\begin{equation}}
\newcommand{\ee}{\end{equation}}

\def\R{{\mathbb R}}

\def\ep{\epsilon}
\def\d{\delta}

\title{Asymptotics of a class of integrals}
\author{A.G. Ramm \\
\small Department of Mathematics\\[-0.8ex]
\small Kansas State University, Manhattan, KS 66506-2602, USA\\
\small \texttt{ramm@math.ksu.edu}}

\date{}
\begin{document}

\maketitle
\begin{abstract} Consider an integral $I(s):=\int_0^T 
e^{-s(x^2-icx)}dx$, where $c>0$  and $T>0$
are arbitrary positive constants.
It is proved that $I(s)\sim \frac{i}{sc}$ as $s\to +\infty$. 
The asymptotic behavior of the integral
$J(s):=\int_0^Te^{s(x^2+icx)}dx$ is also derived.
One has $J(s)\sim \frac{ e^{sT^2+iscT}}{s(2T+ic)}$
as $s\to +\infty$.
\end{abstract}
{\it MSC}: \,\, 41A60

\noindent\textbf{Key words:} asymptotics of integrals;  complex 
phase function.

\section{Formulation of the Result}
Let $c>0$ be a constant and $s>0$ be a large parameter. Consider first the 
integral
\be\label{e1}
I(s):=\int_0^T e^{-s(x^2-icx)}dx,
\ee
where $T>0$ is an arbitrary fixed constant. Our result concerning the 
asymptotics of the integral $I(s)$ is formulated in Theorem 1, below. The 
choice of $T>0$
{\it does not influence} the asymptotics of $I(s)$ as $s\to \infty$. This 
is proved in Remark 1 below. However, the choice of $T$ {\it does 
influence} 
the asymptotics of the integral 
\be\label{e1a}
J(s):=\int_0^T e^{s(x^2+icx)}dx,
\ee
as $s\to \infty$. The asymptotic behavior of the integral $J(s)$ is given 
in 
Theorem 2. This theorem is formulated below and derived in 
Section 3.
In Section 1  one assumes without loss of generality that  $T=\infty$ and 
uses
some analytical results from \cite{BE} and \cite{L}.

The aim of this paper is to derive asymptotic formulas for the 
integrals
$I(s)$ and $J(s)$ as $s\to +\infty$. 
In what follows we write $\infty$ for
$+\infty$. 

Let us formulate our first result.

{\bf Theorem 1.} {\it One has
\be\label{e2}
I(s)\sim \frac{i}{sc}\quad  as\quad s\to +\infty.
\ee
} 
In Section 2 a proof of Theorem 1 is given. 

In Section 3 the
asymptotic behavior of $J(s)$ is found.
The following theorem is proved there.

{\bf Theorem 2.} {\it One has
\be\label{e2a}
J(s)\sim \frac{ e^{sT^2+iscT}}{s(2T+ic)} \qquad  as\quad 
s\to +\infty.
\ee
}
In the large literature on the stationary phase and the steepest 
descent methods the integrals of the type $\int_D e^{i\lambda 
S(x)} f(x)dx$ are studied when $\lambda\to \infty$. It is 
usually
assumed that the phase  $S(x)$ is a real-valued or a purely 
imaginary function, that $D\subset \R^n$ is a bounded domain or the whole 
space $\mathbb{R}^n$, and that the phase function $S(x)$ has finitely many 
non-degenerate critical points in $D$, see, for example, \cite{BH}, 
\cite{F}. 

There are some works on the steepest descent method  when the phase 
function $S(x)$ is assumed analytic and one chooses a steepest descent 
contour on which the phase function is either a real-valued or a 
purely imaginary function. 

We study
two examples in which the phase function is complex-valued. The special
choice of the phase function is motivated by a study of the Pompeiu 
problem,  see
\cite{R470}, Chapter 11. 

Our derivations of the asymptotic formulas for 
the integrals  
$I(s)$ and $J(s)$ differ from the
derivations in \cite{BH} and \cite{F}. Namely, we use some formulas from 
the theory of special functions, see \cite{BE}, \cite{L}, in the proof
of Theorem 1, and some
simple argument in the proof of Theorem 2.
The principally novel point in our results is the fact that the 
phase functions $x^2+icx$ and $-x^2+icx$ are neither real-valued nor 
purely imaginary.

Our presentation is independent
of the published works and has practically no intersection with the 
published literature.

\section{Proof of Theorem 1} 

Let us start with two formulas from \cite{BE}, formulas 1.4.11 and 2.4.18:
\be\label{e3}
\int_0^\infty e^{-ax^2}\cos(xy)dx=\frac {\pi^{1/2}}{2a^{1/2}}e^{-\frac 
{y^2}{4a}}, \qquad a>0,
\ee 
and 
\be\label{e4}
\int_0^\infty e^{-ax^2}\sin(xy)dx=\frac{ye^{-\frac{y^2}{4a}}}{2a} 
F(1/2;3/2;\frac{y^2}{4a}), \qquad a>0, 
\ee
where $F(b;c;x)$ is the degenerate hypergeometric function, defined in
in \cite{L}, Chapter 9, by the formula 
$$F(b;c;x)=\sum_{k=0}^\infty \frac {(b)_k}{(c)_k}\frac{x^k}{k!},$$
where $(b)_k:=\frac {\Gamma(b+k)}{\Gamma(b)}$, $(b)_0:=1.$ In
\cite{BE} the function $F(b;c;x)$ is denoted sometimes by $_1F_1(b;c;x)$. 

One has (\cite{L}, formula (9.12.8)):
\be\label{e5}
F(b;c;x)\sim \frac {\Gamma(c)}{\Gamma(b)}e^x x^{b-c}[1+O(\frac 1 x)], 
\quad x\to +\infty.
\ee
Let $y=cs$ and $a=s$, $b=1/2$, $c=3/2$, $x=\frac {y^2}{4s}$. It follows 
from formulas \eqref{e3}-\eqref{e5}
after some simple algebraic calculations that
\be\label{e6}
I(s)=\int_0^\infty e^{-sx^2}e^{icsx}dx\sim \frac i{sc}, \quad s\to 
\infty.
\ee
Theorem 1 is proved. \hfill $\Box$

{\bf Remark 1.} {\it If $T>0$ is an arbitrary fixed number, then
\be\label{e7}
\int_0^T e^{-sx^2}e^{icsx}dx\sim  \frac i{sc}, \quad s\to \infty.
\ee
This follows from the esimate
\be\label{e8}
|\int_\epsilon^T e^{-sx^2}e^{icsx}dx|\le O(e^{-s\epsilon^2})=o(s^{-1}),
\quad s\to \infty,
\ee
which holds for any number $\epsilon>0$ and any $T>\epsilon$.}

Note that one can 
calculate some other integrals using the formula for $I(s)$. For example,
\be\label{e9}
I_1(s):=\int_0^T e^{-sx^2}e^{icsx}x^2dx=s^{-2} (-id/dc)^2I(s)\sim 
-2is^{-3}c^{-3}, \quad s\to \infty.
\ee
where the differentiation with respect to parameter $c$ is justified.

One may take $T=\infty$ without loss of the generality, as was mentioned 
above, see  Remark 1. 

The result \eqref{e6} can be formally obtained
if one neglects the term $-sx^2$ in the phase, uses standard
asymptotics of the integral $\int_0^a e^{icsx}f(x)dx$ as $s\to 
\infty$,
and assumes that $f\in C^1([0,a])$, $f=1$ in a neighborhood of the origin 
$x=0$ and $f=0$ for $x\ge a$. This is a formal argument, and 
the rigorous justification of the resulting asymptotic formulas (8) and 
(9) is given in the proof of Theorem 1 and in Remark 1.

\section{ Asymptotics of $J(s)$}

Consider now  integral \eqref{e1a}. The principal difference of this 
integral compared with
the integral \eqref{e1} is that the asymptotic behavior of $J(s)$ depends 
now on both parameters $T$ and $c$. Moreover, 
the asymptotical formula for $J(s)$ has a different nature: the 
asymptotic grows exponentially 
and oscillates as $s\to \infty$.

Let us describe the idea of our argument.
The idea is to transform the integrand $e^{sx^2+iscx}$ into 
a form for which the asymptotic behavior can be easily calculated.
This is done as follows.
One has:
\be\label{e10}
J(s)=e^{\frac {sc^2}4}\int_0^T e^{s(x+\frac {ic}2)^2}dx.
\ee
Thus,
\be\label{e10'}
J(s)=e^{\frac {sc^2}4+s(T+\frac {ic}2)^2}\int_0^T e^{s[(x+\frac {ic}2)^2-
(T+\frac {ic}2)^2]}dx.
\ee
After some algebraic transformations one gets
\be\label{e11}
J(s)=e^{sT^2+iscT}\int_0^T e^{-sy(2T+ic-y)}dy, \quad y=T-x.
\ee
Denote
\be\label{e12}
J_1(s):=\int_0^T e^{-sy(2T+ic-y)}dy.
\ee
We derive below, in Lemma 1, the asymptotic formula 
\be\label{e13}
J_1(s)\sim \frac 1{s(2T+ic)},  \qquad s\to \infty.
\ee
Combining formulas \eqref{e11}-\eqref{e13}, one  gets
\be\label{e14}
J(s)\sim \frac {e^{sT^2+iscT}}{s(2T+ic)}, \qquad s\to \infty.
\ee

{\bf Theorem 2.} {\it The asymptotics of $J(s)$ as $s\to \infty$ is given
by formula \eqref{e14}.}

Let us now derive formula \eqref{e13}.  

{\bf Lemma 1.} {\it Formula \eqref{e13} holds.}

{\it Proof.} Let $\ep>0$ be a small number which is specified below.
One has 
\be\label{e15}
J_1(s)=\int_0^{\ep}e^{-sy(2T+ic-y)}dy+O(e^{-s\ep(2T-\ep)}).
\ee  
Here we have used the fact that the function $y(2T-y)$ is monotonically
growing on the interval $[0, T]$. 

Let us calculate the asymptotics of the integral 
$$J_2(s):=\int_0^{\ep}e^{-sy(2T+ic-y)}dy, \qquad s\to \infty.$$
Let us choose $\ep=\ep(s)$ such that $s\ep\to \infty$ while $s\ep^2
\to 0$ as $s\to \infty$. This can be done in many ways. For example,
one may take $\ep=s^{-(0.5+\d)}$, where $\d\in (0, 0.5)$.
With such a choice of $\ep(s)$ one has
\be\label{e16}
J_2(s)=\int_0^{\ep}e^{-sy(2T+ic-y)}dy=\int_0^{\ep}e^{-sy(2T+ic)}dy 
[1+o(1)]\qquad s\to \infty.
\ee
The asymptotics of the integral 
$$J_3(s):=\int_0^{\ep}e^{-sy(2T+ic)}dy$$
can be calculated easily:
\be\label{e17}
J_3(s)=\int_0^{\ep}e^{-sy(2T+ic)}dy\sim \frac 1 {s(2T+ic)}, \qquad s\to 
\infty.
\ee
Note that $O(e^{-s\ep(2T-\ep)})=o(s^{-1})$ as $s\to \infty$
and $\ep=\ep(s)$ is chosen, for example,  so that 
$e^{-s\ep(s)T}=o(s^{-1})$. This relation holds, for example, if
$\ep(s)=s^{-0,6}$.
From \eqref{e15}-\eqref{e17} the conclusion of Lemma 1 follows.\hfill $\Box$

\end{document}